# Easy-Plane Alignment of Anisotropic Biofluid Crystals in a Magnetic Field: Implications for Rod Orientation

Robert J. Deissler and Robert Brown

Department of Physics, Case Western Reserve University, Cleveland, Ohio 44106 USA

*Index Terms* – Anisotropic, beta-hematin, crystals, gout, hemozoin, magnetic field, malaria, susceptibility.

We study the orientation in a uniform magnetic field of rod-like anisotropic biofluid crystals with an easy plane that makes an oblique angle with the crystal's c-axis. For a sufficiently strong field, these crystalline rods orient themselves such that the crystal's easy plane is parallel to the magnetic field, the rod's direction being defined as the direction of the crystal's c-axis. As the rod rotates about the crystal's hard axis there will therefore be a range of angles that the rod makes with the magnetic field. We detail this behavior by first providing illustrations of hemozoin crystals at various orientations. These illustrations clearly demonstrate that the orientation angle that the crystalline rod makes with respect to the magnetic field varies from about 30º to 150º. We also derive an analytical expression for the probability density function for the orientation angle. We find that the orientation angles are not uniformly distributed between the limits of 30º and 150º, but rather tend to cluster near these limits. This suggests experimental tests and addresses confusion about the rod orientation found in past literature. The relevance to other anisotropic biofluid crystals, such as those produced by gout, is also discussed.

## I. Introduction

Human disease in important cases may be characterized by crystal formation. The crystals in the notable examples of malaria [1], [2], [3], [4], [5], [6] and gout [7], [8] exhibit magnetic susceptibility anisotropy and linear dichroism, properties that can lead to methods of detection, since an applied magnetic field will cause the crystals to rotate and the rotation in turn changes the absorption of polarized light transmission through the fluid. In the present paper, we detail the interaction of an external magnetic field with anisotropic crystals suspended in a fluid, the key in simple detectors. The orientation of the crystals is of primary focus here.

We first consider the production of hemozoin, the crystals [1], [4], [9], [10], [11] relevant to malaria infections as a specific case to best explain the orientation results, the principal subject of our study. (While some of the studies we reference are of the synthetic version of hemozoin, beta-hematin, the crystal structure is the same. Therefore, our work is relevant to either hemozoin or beta-hematin.) The parasites, injected into the bloodstream by mosquitos, enter red blood cells and go first through a stage of taking in cytoplasm, which is rich in hemoglobin. That hemoglobin is digested into peptides and degraded into amino acids. The parasite must avoid a heme toxicity and hence this sequestration of free heme (iron bound to four nitrogen atoms – a ring of protoporphyrin having two carboxylic side chains). The iron at the center generates free radicals, and hence disrupts protein structure. Within the sequestered organelle, the hemes dimerize to form hemozoin through a bond between the iron of one heme and an oxygen of the carboxylic side chain of a second heme. This bond makes the hemozoin paramagnetic. Consequently, hemozoin forms hydrogen bonds with the oxygen of their second carboxylic side chains. Hemozoin consists of layers of these long chains, held together by π-π stacking forces. The whole hemozoin crystal is paramagnetic as a result, leading to a stronger interaction with an external magnetic field, than diamagnetic uninfected blood.

The influence of an external magnetic field on the crystalline structure is richer than the simple example of a compass needle. While the hemozoin crystals have an approximate rod-like form, the reality is that the rods do not align with a uniform field, nor are they perpendicular to the field [1], [4], [10], [11]. The key is to find the easy plane, which in the hemozoin case is the porphyrin plane. This plane contains an inner ring of nitrogen around iron, an outer ring of carbon and an outside hydrogen. The plane is found generally at an oblique angle with the rod, the direction of the rod being defined as the direction of the c-axis of the crystal. This angle is approximately 30º for hemozoin [1], [4], [10], [11]. In the presence of a sufficiently strong magnetic field, the rods orient themselves such that the magnetic field vector lies in the porphyrin plane, thereby reducing the number of rotational degrees of freedom from three to two, the two degrees of freedom being rotation about the magnetic field vector and rotation about the hard axis, which is perpendicular to the easy plane. We detail this behavior and its modeling for the hemozoin example in the work presented here.

The reaction of a concentration of hemozoin crystals to the presence of a magnetic field is a key to their detection. A particular example is an application in which the absorption of a light beam passing through infected blood changes when the field is turned on and off. Different versions of magneto-optical detection have been proposed and carried out with success in malaria diagnostics [1], [2], [3], [4], [5], [6], [12], [13], [14], [15]. A scenario exhibiting a larger increase in the absorption is a light beam polarized in the same direction as the magnetic field; the polarization is therefore also in the porphyrin plane. The experiments suggest that light interacts most strongly with the electrons in that case. The absorption is reduced in the unpolarized case or, equivalently, the thermally randomized crystal case in which the field is absent. Reference [1] focuses on the details of these cases.

In the present paper, we examine in detail the orientation of rod-like anisotropic crystals suspended in a fluid under the influence of an external uniform magnetic field. The anisotropy arises, as we have discussed, by the existence of an easy plane, fixed relative to the geometrical structure. It was argued in [1] that in a sufficiently strong magnetic field, a hemozoin crystalline rod makes an angle $\theta$ between about 30º and 150º with the magnetic field vector as the crystal rotates about its hard axis due, for example, to thermal motion. We investigate in more depth the picture behind this scenario, providing illustrations of hemozoin crystals at various orientations. At first glance, it may appear that the angle the rod makes with the magnetic field vector would be uniformly distributed between 30º and 150º. However, we find that this is not the case. In the next section we derive an analytical expression for the probability density function (PDF) for the orientation angle that the rod makes with respect to the magnetic field vector. We find that the PDF for this angle is not uniform, but rather increases as $\theta \to \pi/6$ and as $\theta \to 5\pi/6$, causing the orientation angles to cluster near the limits $\theta = \pi/6$ and $\theta = 5\pi/6$. A concluding section addresses general lessons for the orientation in a magnetic field of anisotropic crystals suspended in a fluid.

## II. Theory and results

Figure 1 shows the packing diagram as well as some illustrations of hemozoin crystals. As seen in Fig. 1a, the c-axis makes an angle of about 30º with the porphyrin plane. Corresponding to the packing diagram of Fig. 1a, Fig. 1b illustrates a hemozoin rod which makes an angle of 30º with the porphyrin plane. Assuming the magnetic field is sufficiently strong so that the magnetic field vector lies in the porphyrin plane, Fig. 1b therefore shows a rod which makes an angle of 30º with the **B** field. The crystal may be rotated about **B** or about the hard axis, and **B** will still lie in the porphyrin plane as illustrated in Fig. 1c, which shows a rod rotated by 30º about **B** and 30º about the hard axis.

Figure 2 shows illustrations of the rod being rotated about **B** in steps of 30º from left to right and, alternatively, about the hard axis in steps of 30º from top to bottom. As demonstrated by these illustrations, the rod may be rotated about **B** or about the hard axis, and the B field will still lie in the porphyrin plane and be perpendicular to the hard axis. As the rod is rotated about the hard axis, as illustrated from top to bottom in Fig. 2, the angle the rod makes with the B field varies from about 30º to 150º, as noted earlier. Note that if a light beam is polarized in the direction of the B field, the light beam's electric field will lie in the porphyrin plane regardless of the orientation angles of the rods, which maximizes absorption, as previously mentioned in the Introduction. It should also be noted that in the case of hemozoin, it is only the orientation of the porphyrin planes that matters regarding the absorption of polarized light, the angles of the crystals themselves about the hard axes being irrelevant [1].

We now consider more generally a rod-like anisotropic crystal suspended in a fluid with an easy plane that makes an oblique angle with the c-axis, the rod direction being defined as the direction of the c-axis. We define three angles. Let $\theta$ be the angle that the rod makes with the B field, let $\theta_0$ be the angle that the rod makes with the easy plane, or the minimum angle that the rod makes with the B field (about 30º or $\pi/6$ for hemozoin), and let $\xi$ be the angle of the crystal about its hard axis. We can relate these angles by

$$\cos\theta = \cos\theta_0 \cos\xi \qquad (1)$$

or

$$\theta = \cos^{-1}(\cos\theta_0 \cos\xi) . \qquad (2)$$

Figure 3 shows a plot of $\theta$ as a function of $\xi$ for fixed $\theta_0 = \pi/6$, the angle between the c-axis and the easy plane. The angle $\theta$ has the range $\pi/6$ to $5\pi/6$ as expected. As $\xi$ varies from 0 to $2\pi$, $\theta$ varies from $\pi/6$ to $5\pi/6$ and back to $\pi/6$.

Assuming that the PDF is uniform for $\xi$ – that is, any angle about the hard axis is equally likely to occur – we may calculate the PDF for $\theta$ by using the change of variable formula [16]

$$g(\theta) = f(\xi)\left|\frac{d\xi}{d\theta}\right|, \tag{3}$$

where $f$ is the PDF for the angle $\xi$ about the hard axis, and $g$ is the PDF for the orientation angle $\theta$ that the rod makes with the B field.

Referring to Fig. 3 and noting that the full range of $\theta$ from $\theta_0$ to $\pi - \theta_0$ is obtained by taking $0 < \xi < \pi$, the normalized PDF for $\xi$ on this interval is given by the constant $f(\xi) = 1/\pi$, which is uniform as noted and normalized to unity. Taking $d/d\theta$ of both sides of Eq. (1), the PDF for the angle the rod makes with respect to the B field is found from (3) to be

$$g(\theta) = \frac{1}{\pi} \frac{\sin\theta}{\sqrt{\cos^2\theta_0 - \cos^2\theta}}. \tag{4}$$

It is easy to show that this function is also normalized to unity on the interval $\theta_0 < \theta < \pi - \theta_0$. To verify this, make the change of variables $x = \cos\theta$ and define $x_0 = \cos\theta_0$; we have

$$\int_{\theta_0}^{\pi-\theta_0} g(\theta)d\theta = \frac{1}{\pi} \int_{-x_0}^{x_0} \frac{dx}{\sqrt{x_0^2 - x^2}} = 1, \tag{5}$$

where the definite integral can be easily done by changing variables again, or one can use the Mathematica online integral calculator [17], with an eye toward the indefinite integral coming up.

Figure 4 shows a plot of $g(\theta)$ from Eq. (4) for $\theta_0 = \pi/6$. Since the PDF approaches $\infty$ at $\theta = \pi/6$ or at $\theta = 5\pi/6$, the orientation angles would tend to cluster near these limits. To see this clustering it is necessary to divide the interval from $\theta = \theta_0$ to $\theta = \pi - \theta_0$ into bins and calculate the probability for the orientation angles to lie in the various bins. The probability for the orientation angle to lie between $\theta_1$ and $\theta_2$ is given by

$$P(\theta_1 < \theta < \theta_2) = \int_{\theta_1}^{\theta_2} g(\theta)d\theta, \tag{6}$$

where $\theta_0 < \theta_1 < \theta < \theta_2 < \pi - \theta_0$. Using the expression for the indefinite integral (again doing by hand or by [17])

$$\int \frac{dx}{\sqrt{x_0^2 - x^2}} = \tan^{-1}\left(\frac{x}{\sqrt{x_0^2 - x^2}}\right) + \text{constant}, \tag{7}$$

we find

$$P(\theta_1 < \theta < \theta_2) = \frac{1}{\pi}\tan^{-1}\left(\frac{\cos\theta_1}{\sqrt{\cos^2\theta_0 - \cos^2\theta_1}}\right) - \frac{1}{\pi}\tan^{-1}\left(\frac{\cos\theta_2}{\sqrt{\cos^2\theta_0 - \cos^2\theta_2}}\right) \tag{8}$$

for the probability that the orientation angle $\theta$ lies between $\theta_1$ and $\theta_2$. Dividing the interval from $\theta = \theta_0$ to $\theta = \pi - \theta_0$ into bins, Eq. (8) is used to calculate the probability that the orientation angle lies within a given bin. Figure 5 shows the probability distribution for 8 bins and 32 bins for $\theta_0 = \pi/6$. Even for only 8 bins, the probability of the orientation angle being binned adjacent to $\pi/6$ or $5\pi/6$ is twice that of being binned near $\pi/2$. For 32 bins this factor is 3.7.

We expect that this ratio of the probability of the orientation angle being binned adjacent to $\theta = \theta_0$ or $\theta = \pi - \theta_0$ to that of being binned near $\pi/2$ will approach infinity as the number of bins $N$ approach infinity, but an interesting question is how it approaches infinity. Define $\delta\theta$ as the width of a bin. Then the probability of $\theta$ being between $\theta_0$ and $\theta_0 + \delta\theta$ is given by $P(\theta_0 < \theta < \theta_0 + \delta\theta)$ as determined by Eq. ((8). Similarly, the probability of $\theta$ being between $\pi/2 - \delta\theta$ and $\pi/2$ is given by $P(\pi/2 - \delta\theta < \theta < \pi/2)$. Now define the ratio of these two probabilities by

$$R(\delta\theta; \theta_0) \equiv \frac{P(\theta_0 < \theta < \theta_0 + \delta\theta)}{P(\pi/2 - \delta\theta < \theta < \pi/2)}, \qquad (9)$$

where the probabilities $P$ are determined from Eq. (8).

In the limit of small $\delta\theta$ (or large $N$) we have

$$R(\delta\theta; \theta_0) \simeq \sqrt{\frac{2\sin\theta_0 \cos\theta_0}{\delta\theta}} = \sqrt{\frac{2\sin\theta_0 \cos\theta_0}{\pi - 2\theta_0} N}, \qquad (10)$$

where we used the asymptotic relationship $\tan^{-1} x \simeq \pi/2 - 1/x$ for large $x$ to evaluate the probability in the numerator of Eq. (9) and used the relationship $\delta\theta = (\pi - 2\theta_0)/N$.

Figure 6 shows a plot of the probability ratio $R(\delta\theta; \theta_0)$ for $\theta_0 = \pi/6$ as a function of the number of bins $N$ on a log-log plot for both a discreet number of bins as evaluated from Eq. (8) and (9), as well as for a range of $N$ as determined from the approximation Eq. (10). For $N$ greater than about 200 the agreement between the exact and approximate probability ratios is seen to be excellent. Also, since $R \propto \sqrt{N}$ from Eq. (10), we expect that the slope of the log-log plot would be ½, which is easy to verify from Fig. 6.

## III. Conclusions

In this paper we studied the orientation in a uniform magnetic field of a suspension of rod-like anisotropic crystals with an easy plane that makes an oblique angle with the c-axis of the crystal. At the outset we wanted to address a confusion that can arise about the orientation of the crystalline rod. The rod itself is not aligned parallel nor is it perpendicular to the field direction. Rather, in a sufficiently strong field the crystals orient themselves such that the crystal's easy plane is parallel to the magnetic field. Therefore, as the rod rotates about the crystal's hard axis there will be a range of angles that the rod can make with the magnetic field vector. We detail this behavior by first providing illustrations of hemozoin crystals at various orientations. These illustrations clearly demonstrate that the orientation angles that the hemozoin rods make with respect to the magnetic field vector vary from about 30º to 150º. We also derived an analytical expession for the probability density function for these orientation angles. We find that the orientation angles are not uniformly distributed between the limits of 30º and 150º, but rather tend to cluster near these limits. This clustering was demonstrated by dividing the interval from 30º to 150º into a number of bins and plotting the probability distribution. We found that the probability of the orientation angle being binned adjacent to 30º or 150º is in fact larger than that of being binned near 90º, this probability ratio $R$ increasing with the number of bins $N$ and asymptotically being proportional to $\sqrt{N}$ in the limit of large $N$.

We propose experimentally testing the theoretical behavior predicted here. For example, do the orientation angles with respect to the B field tend to cluster near the limits of 30º and 150º for hemozoin? Does the probablilty distribution agree with the the theoretical analysis provided here? If there is an experimental discrepancy, the assumptions about the dominant induced magnetice dipole moments may need revisiting, for instance. Also, it would be interesting to study another biofluid crystal case such as gout [7], [8]. In gout, the body produces monosodium urate (MSU) crystals in the synovial fluid in joints. In [7] it was found that the ring surface for an MSU crystal makes an angle of about 70º with the crystalline rod. Based on the analysis presented here we would expect the orientation angles to vary from about 70º to 110º in the presence of a strong B field, assuming that the plane of the ring surface is indeed an easy plane. Although there is some indication from the experiments in [7] that the orientation angles cluster near 70º and 110º, this may be caused by these large crystals – on the order of tens of microns in length – settling to the bottom of the cell, thus restricting rotation about the hard axis and limiting the range of orientation angles available to the crystals. Therefore, further experimentation is required.

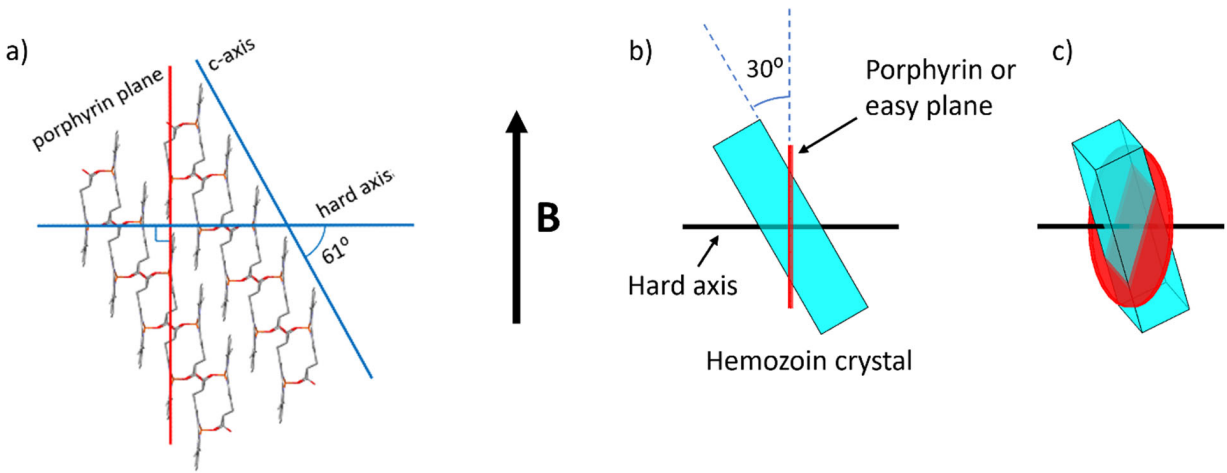

**Fig. 1.** a) Packing diagram of hemozoin showing its crystalline structure. The c-axis is seen to make an angle of about 30⁰ with the porphyrin or easy plane, the hard axis being perpendicular to the porphyrin plane. Reprinted from [1] rotated by 29⁰. b) The crystals are shaped like rods; the direction of a rod being defined as the direction of the c-axis of the crystal. The rods orient themselves such that the B field lies in the porphyrin (or easy) plane, and perpendicular to the hard axis. c) The rod may be rotated about **B** or about the hard axis, and **B** will still lie in the easy plane and be perpendicular to the hard axis as illustrated here, where the rod is rotated 30⁰ about **B** and 30⁰ about the hard axis. Although hemozoin crystals are triclinic, and therefore the edges of a crystal do not form right angles with one another, here we are mainly interested in the orientation of the c-axis of the crystal and for simplicity illustrate the shape as rectangular.

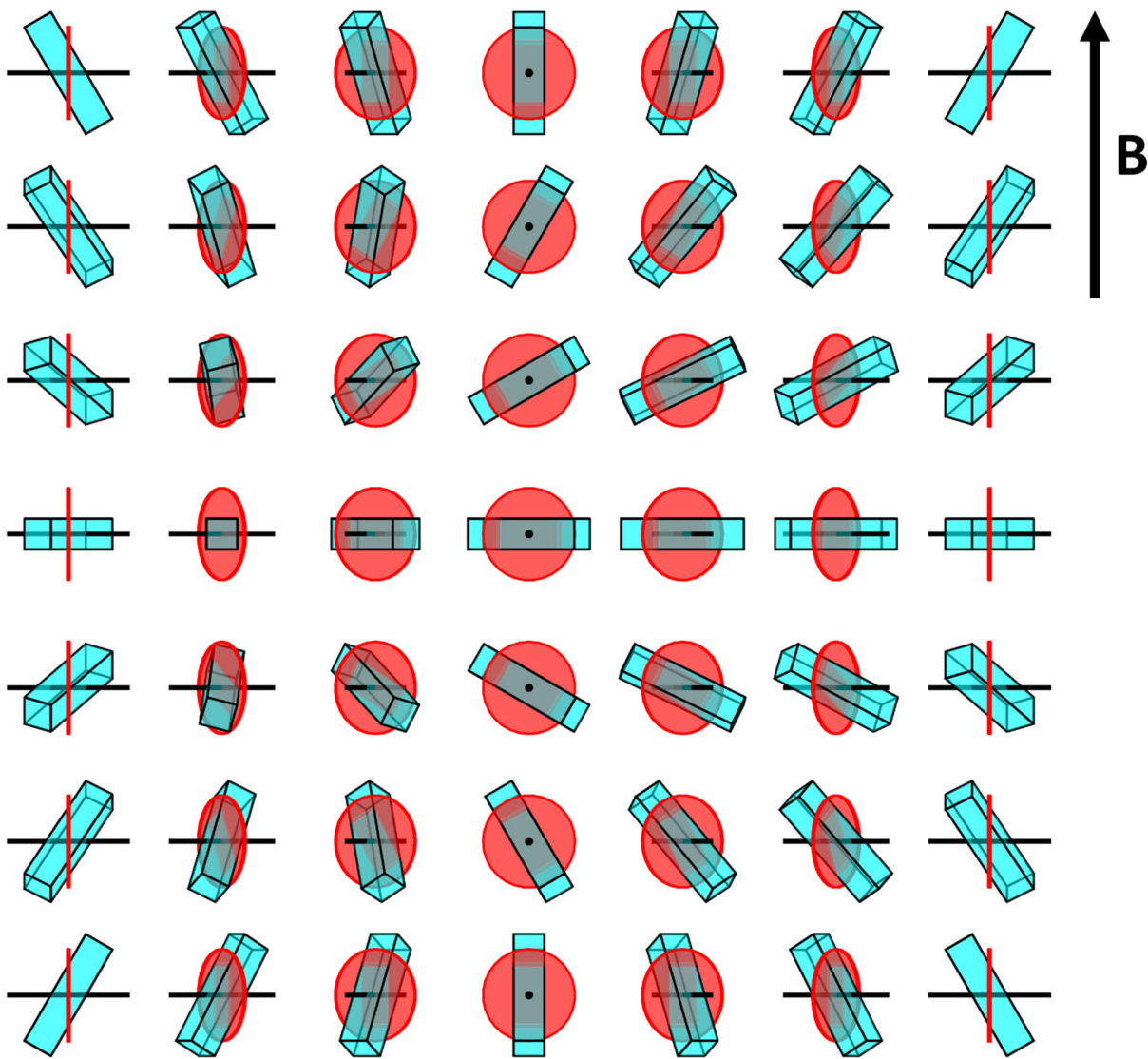

**Fig. 2.** The hemozoin crystal may be rotated about the B field vector or about the hard axis and **B** will still lie in the porphyrin plane as illustrated above. From left to right, the crystalline rod is rotated about **B** in steps of 30⁰. From top to bottom, the rod is rotated about the hard axis in steps of 30⁰. Rotation about the hard axis corresponds to the rod orientation angle being between 30⁰ and 150⁰ with respect to **B**.

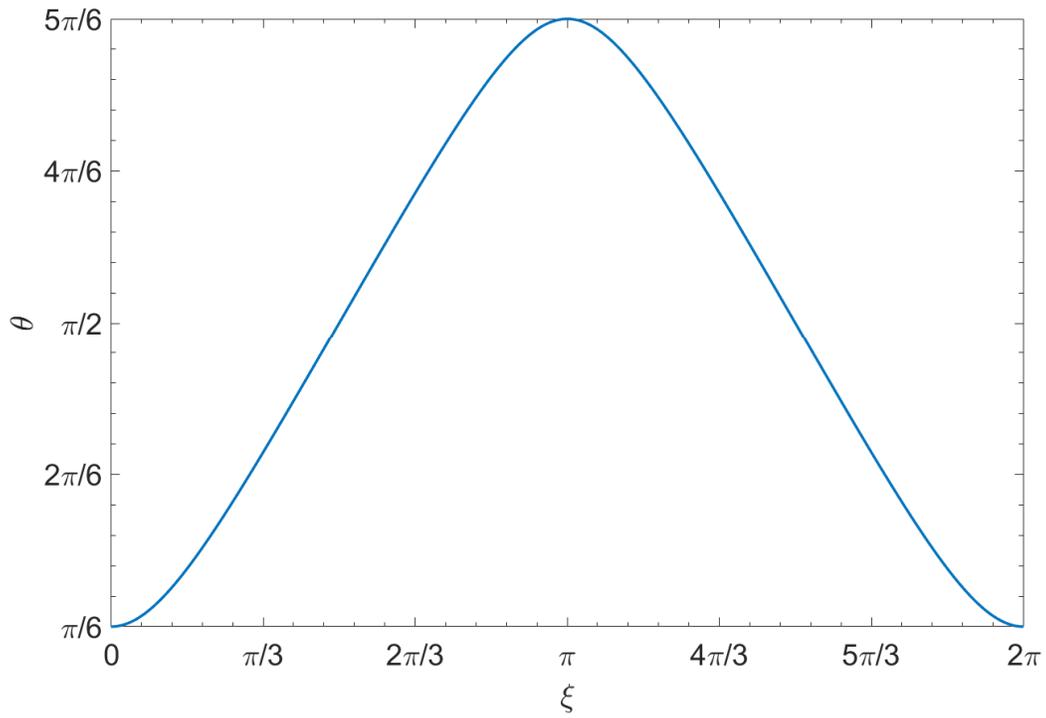

**Fig. 3.** Plot of $\theta$ as a function of $\xi$ from Eq. (2) for $\theta_0 = \pi/6$, which corresponds to hemozoin. As expected, the angle $\theta$ varies from $\pi/6$ to $5\pi/6$.

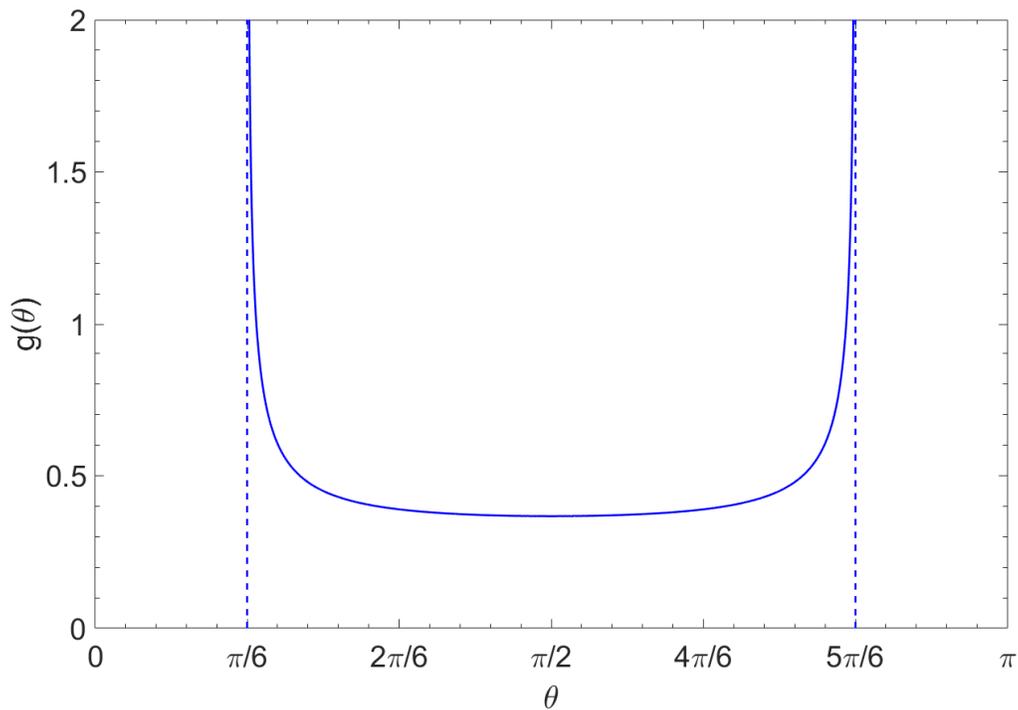

**Fig. 4.** Probability density function for the orientation angle $\theta$ from Eq. (4) for $\theta_0 = \pi/6$, which corresponds to hemozoin. For an ensemble of crystals, their orientation angles will tend to cluster near $\theta = \pi/6$ and $\theta = 5\pi/6$. The asymptotes are given by the dotted lines.

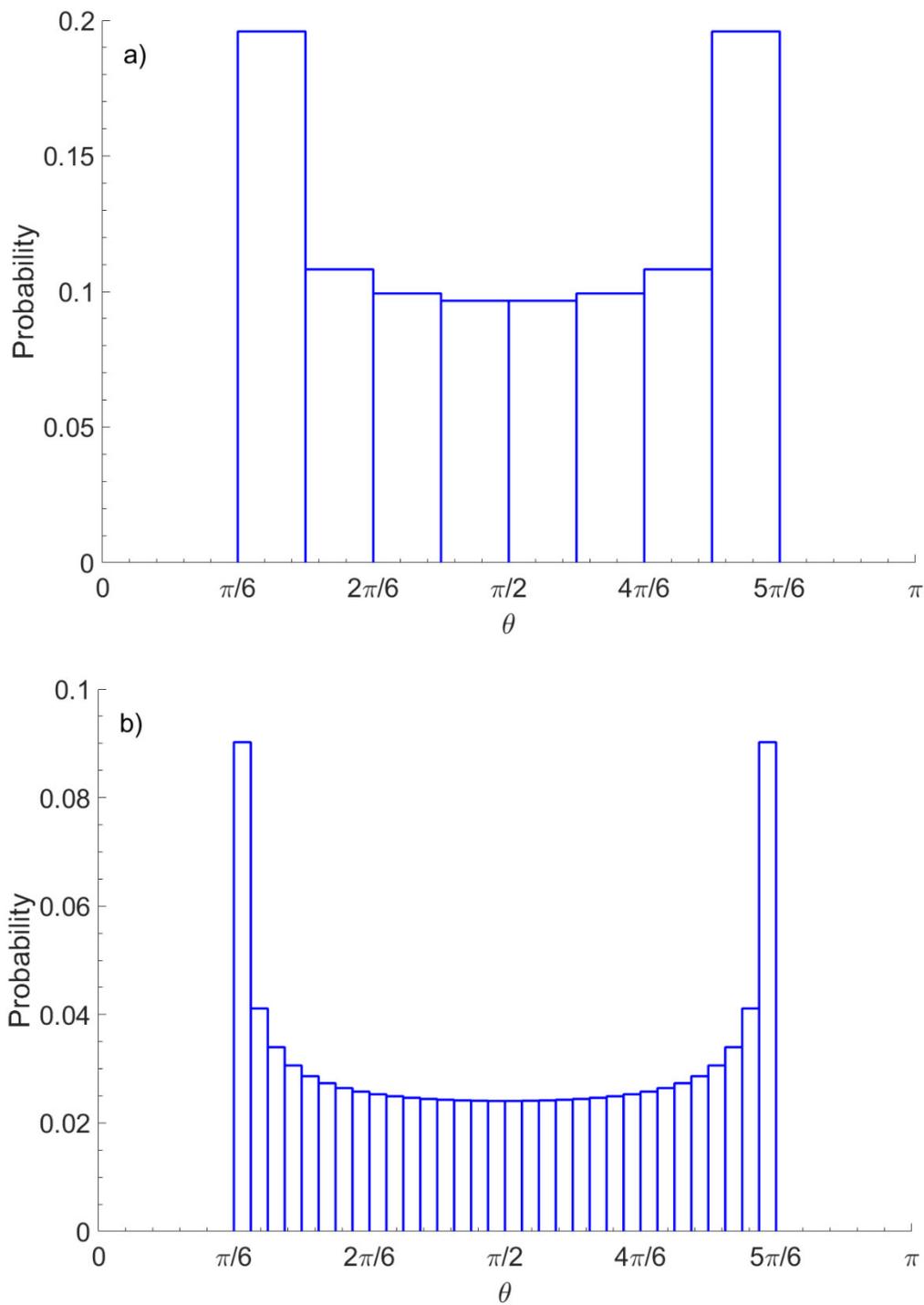

**Fig. 5.** Probability distribution for the orientation angle $\theta$ for $\theta_0 = \pi/6$, which corresponds to hemozoin, for a) 8 bins and b) 32 bins. Even for only 8 bins, the probability of the orientation angle being binned adjacent to $\pi/6$ or $5\pi/6$ is twice that of being binned near $\pi/2$. For 32 bins this factor is 3.7.

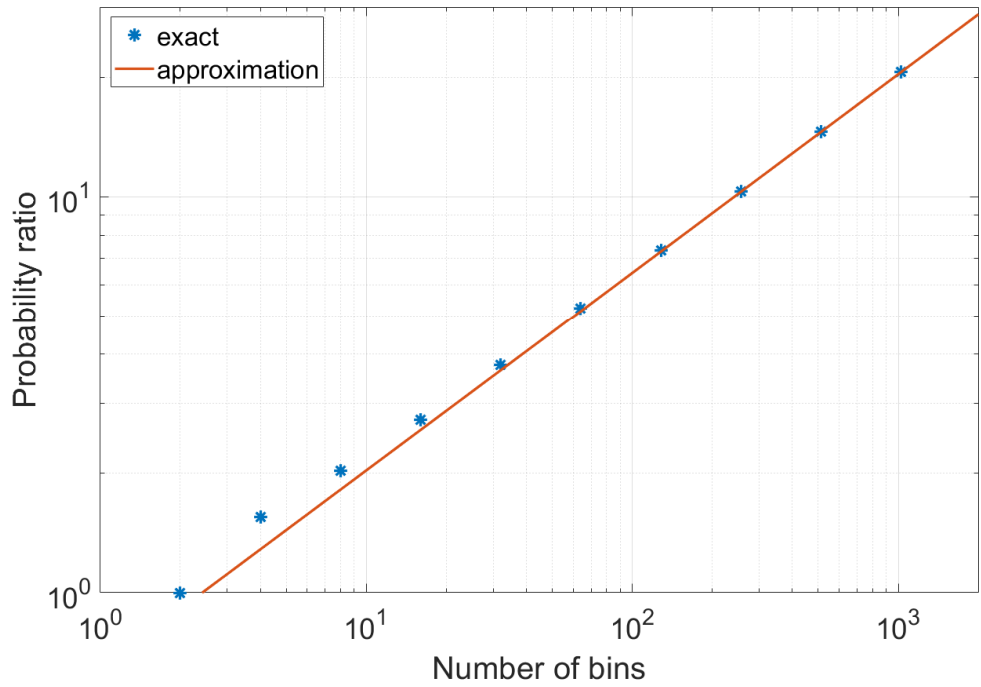

**Fig. 6.** Ratio of the probability of the orientation angle being binned adjacent to $\pi/6$ or $5\pi/6$ to that of being binned near $\pi/2$ as a function of the number of bins on a log-log plot. The exact probability ratio $R$ is evaluated from Eq. (8) and (9) for a number of discreet bin numbers $N$, and the approximate probability ratio is determined from Eq. (10), with $\theta_0 = \pi/6$, which corresponds to hemozoin.